\documentclass[sn-vancouver,iicol,Numbered]{sn-jnl} 

\usepackage{xspace}
\usepackage{setspace}
\usepackage{amssymb}

\usepackage{amsmath}
\usepackage{multirow}
\usepackage{gensymb}
\usepackage{mathtools}

\usepackage{adjustbox,lipsum,bm,soul}

\usepackage{tikz}
\usepackage{gensymb} 

\usepackage{amsfonts,amsmath,amsthm,amssymb,graphicx}

\usepackage{color}
\usepackage{longtable}

\usepackage{algorithm}%
\usepackage{algorithmicx}%
\usepackage{algpseudocode}%

\newcommand{\discus}{DISCUS\xspace}

\renewcommand{\eqref}[1]{Equation~(\ref{eq:#1})}

\newcommand{\tabref}[1]{Table~\ref{tab:#1}}
\newcommand{\figref}[1]{Figure~\ref{fig:#1}} 

\DeclareMathOperator*{\argmin}{arg\,min}

\renewcommand{\Hat}{\widehat}

\renewcommand{\vec}[1]{\ensuremath{\boldsymbol{#1}}}

\newcommand{\hvec}[1]{\ensuremath{\Hat{\boldsymbol{#1}}}}

\newcommand{\Real}{{\mathbb{R}}}
\newcommand{\Complex}{{\mathbb{C}}}

\usepackage{marginnote}

\usepackage{caption}
\usepackage{newfloat}

\DeclareFloatingEnvironment[
    fileext=los,
    listname={List of Supplementary Figures},
    name=Figure,
    placement=htbp,
    within=none
]{suppfigure}

\DeclareCaptionLabelFormat{suppfig}{Figure S#2~}
\newcommand{\textr}[1]{\textcolor{black}{#1}}
\newcommand{\textb}[1]{\textcolor{black}{#1}}
\begin{document}
\title[Article Title]{An unsupervised method for MRI recovery: Deep image prior with structured sparsity}

\author[1]{\fnm{Muhammad Ahmad} \sur{Sultan}}\email{sultan.47@osu.edu}
\author[1,2]{\fnm{Chong} \sur{Chen}}\email{chong.chen@osumc.edu}
\author[3]{\fnm{Yingmin} \sur{Liu}}\email{yingmin.liu@osumc.edu}
\author[4]{\fnm{Katarzyna} \sur{Gil}}\email{katarzyna.gil@osumc.edu}
\author[4]{\fnm{Karolina} \sur{Zareba}}\email{karolina.zareba@osumc.edu}
\author*[1,2,3]{\fnm{Rizwan} \sur{Ahmad}}\email{ahmad.46@osu.edu}

\affil[1]{\orgdiv{Biomedical Engineering}, \orgname{Ohio State University}, \orgaddress{\city{Columbus}, \state{OH} \postcode{43210}, \country{USA}}}
\affil[2]{\orgdiv{Electrical \& Computer Engineering}, \orgname{The Ohio State University}, \orgaddress{\city{Columbus}, \state{Ohio} \postcode{43210}, \country{USA}}}
\affil[3]{\orgdiv{Davis Heart and Lung Research Institute}, \orgname{The Ohio State University Wexner Medical Center}, \orgaddress{\city{Columbus}, \state{Ohio} \postcode{43210}, \country{USA}}}
\affil[4]{\orgdiv{Division of Cardiovascular Medicine}, \orgname{The Ohio State University Wexner Medical Center}, \orgaddress{\city{Columbus}, \state{Ohio} \postcode{43210}, \country{USA}}}

\abstract{
\textbf{Objective:} To propose and validate an unsupervised MRI reconstruction method that does not require fully sampled k-space data.
\vspace{1mm}

\textbf{Materials and Methods:} The proposed method, deep image prior with structured sparsity (\discus), extends the deep image prior (DIP) by introducing group sparsity to frame-specific code vectors, enabling the discovery of a low-dimensional manifold for capturing temporal variations. \discus was validated using four studies: (I) simulation of a dynamic Shepp-Logan phantom to demonstrate its manifold discovery capabilities, (II) comparison with compressed sensing and DIP-based methods using simulated single-shot late gadolinium enhancement (LGE) image series from six distinct digital cardiac phantoms in terms of normalized mean square error (NMSE) and structural similarity index measure (SSIM), (III) evaluation on retrospectively undersampled single-shot LGE data from eight patients, and (IV) evaluation on prospectively undersampled single-shot LGE data from eight patients, assessed via blind scoring from two expert readers.
\vspace{1mm}

\textbf{Results:} \discus outperformed competing methods, demonstrating superior reconstruction quality in terms of NMSE and SSIM (Studies I--III) and expert reader scoring (Study IV).
\vspace{1mm}

\textbf{Discussion:} An unsupervised image reconstruction method is presented and validated on simulated and measured data. These developments can benefit applications where acquiring fully sampled data is challenging.}

\maketitle
\section{Introduction}\label{sec:intro}
Magnetic resonance imaging (MRI) is a versatile imaging modality that offers excellent soft-tissue contrast and high spatial and temporal resolution, all without using ionizing radiation. MRI is routinely used in a broad range of clinical applications, including neuro, musculoskeletal, abdominal, and cardiovascular imaging. Cardiovascular MRI is considered the gold standard for measuring cardiac function using cine and for assessing myocardial scar using late gadolinium enhancement (LGE). Despite its clinical significance, cardiovascular MRI suffers from long scan times, resulting in patient discomfort, decreased throughput, and increased susceptibility to motion artifacts. This limitation is often addressed by prospective undersampling of k-space to complete the scan faster. To deal with the resulting ill-posed inverse problem, several image reconstruction approaches have been proposed \cite{ravishankar2017acceleratingMRI}.
Most commercial scanners employ reconstruction methods that leverage both parallel imaging, where data are simultaneously acquired across multiple receive coils \cite{pruessmann1999sense, griswold2002GRAPPA, uecker2014espirit}, and compressed sensing (CS) techniques, which utilize explicit sparsity-based priors \cite{lustig2007CS, otazo2015LplusS}. However, the acceleration and image quality offered by these techniques are limited, and speeding up scans and improving image quality remains an ongoing challenge.

More recently, deep learning (DL) methods have shown great promise in further accelerating MRI. These methods have been shown to outperform the traditional CS and parallel imaging methods in terms of image quality at higher acceleration rates \cite{zbontar2018fastmri}. Some supervised DL methods pose the image recovery problem as an end-to-end dealiasing problem where a convolutional neural network (CNN) takes in the coil-combined aliased image and generates a fully formed image, without any guidance from MRI physics \cite{hyun2018DL}. However, such approaches typically do not generalize well when the forward model varies between training and testing stages. Moreover, they require extensive training data, which are not generally available for dynamic MRI \cite{chen2020ocmr}. In contrast, other supervised DL methods incorporate guidance from the MRI physics in the training process by explicitly performing data consistency within an unrolled network \cite{hammernik2018VarNetFastMRI,aggarwal2018modl}. These methods are generally more robust and offer state-of-the-art performance but still require fully sampled data for training. Plug-and-play methods provide yet another option for image reconstruction, where an off-the-shelf generic denoiser or an application-specific DL-based denoiser is repeatedly called within an iterative algorithm \cite{venkatakrishnan2013PnP,ahmad2020plug}. These methods do not require fully sampled k-space data as the denoiser can be trained on high-quality image patches. However, such image patches are not available for many MRI applications, including cardiac imaging.

To address the limited availability of the training data, several self-supervised and unsupervised DL methods have been proposed for MRI reconstruction, including deep image prior (DIP) and deep decoder \cite{ulyanov2018DIP,heckel2018deep}. These methods model an image as the output of a generator network, with both network parameters and input latent code vectors trained on an image-specific basis. DIP-based methods utilize the CNN network structure as an implicit prior \cite{chakrabarty2019DIPbias}. Since they are prone to overfitting, early stopping is often required for applications where measured data are noisy \cite{wang2021early}. More recently, Bell et al. proposed a noise-robust extension of DIP by training the self-guided network to function as a denoiser instead of a generator and applied it to knee MRI \cite{bell2023SG_DIP}. 
\textr{Recent review articles by Hammernik et al. \cite{hammernik2023physics_DL_review} and Heckel et al. \cite{heckel2024DL_review} provide a comprehensive summary of state-of-the-art deep learning methods for MRI reconstruction.}

For dynamic MRI, various extensions of DIP have been proposed to recover a series of images. Yoo et al. employed DIP by training a generative network to map a low-dimensional manifold to a cine image series \cite{yoo2021TD_DIP}. Their work, however, requires pre-estimating the approximate number of cardiac cycles from the radial data and relies on smooth variations in latent space to capture continuous dynamics in the cardiac cine series. At the same time, Zou et al. independently proposed a similar method, called Gen-SToRM \cite{zou2021Gen_SToRM}, which uses spiral sampling and explicitly enforces smoothness on the latent code vectors to recover cine images. These methods define the manifold dimensionality in advance and rely on the temporal smoothness of latent code vectors. In general, the manifold dimensionality is not known precisely due to the multiple unknown sources of motion and contrast changes in cardiac MRI. Moreover, for single-shot applications, where the data are not collected continuously, the latent code vectors are not expected to be temporally smooth. More recently, Ahmed et al. \cite{ahmed2022deblur} and Hamilton et al. \cite{hamilton2023LR_DIP} integrated low-rank constraint with DIP to facilitate accelerated free-breathing cardiac cine imaging. 
In a related line of work, Zou et al. trained the network to generate frame-specific 3D deformation fields for free-breathing lung imaging \cite{zou2022dynamic}. These deformation fields can generate an image series by deforming a single template image, which is also jointly estimated with the deformation fields. This method is more appropriate for 3D imaging where there are no temporal changes in the image content other than the non-rigid motion. 

\textr{Apart from manifold learning, image registration-based methods have also been explored as a post-reconstruction step or to guide the reconstruction. For example, Morales et al. \cite{morales2019pairwise_deformations} proposed a 3D CNN to generate pairwise motion fields to register 2D cine images. Lu et al. \cite{lu2023bidirectional_groupwise} introduced an unsupervised registration-based motion tracking model for 2D cine images. Yang et al. \cite{yang2022groupwise_dynamic} integrated groupwise motion field estimation into a unified deep learning framework for 2D cine reconstruction. Ghoul et al. \cite{ghoul2024attention_pairwise} enabled motion-compensated accelerated imaging by leveraging an attention mechanism to extract local and global features for pairwise motion field estimation.}

In this work, we propose an extension of DIP, called deep image prior with structured sparsity (\discus). \discus trains a single network to map a series of random code vectors to a series of images. In contrast to methods that require specifying the dimensionality of the manifold \cite{zou2021Gen_SToRM}, which may not be known in advance, \discus discovers the dimensionality by imposing group sparsity on the dynamic code vectors. Additionally, \discus does not assume that temporal closeness (order of acquisition) is tied to image similarity \cite{zou2021Gen_SToRM,yoo2021TD_DIP}. This makes \discus suitable for single-shot applications, e.g., free-breathing LGE imaging and parametric mapping, where consecutive frames are not necessarily more similar. 

This work builds upon our preliminary results \cite{sultan2024discus} and now includes an additional simulation study, an ablation study, comparison with state-of-the-art methods, and an application to prospectively undersampled LGE data. In the subsequent sections, we describe \discus in detail and present four different numerical studies for its evaluation and validation.

\section{Materials and Methods}\label{sec:mnm}
\subsection{DIP}

In MRI, the complex-valued data are measured in the spatial frequency domain, called k-space. The reconstruction process involves estimating the underlying image from noisy and potentially undersampled k-space measurements. The measured noisy data are related to the image by
\begin{align}
\scalebox{0.8}{%
$\begin{aligned}
\label{eq:model}
\vec{y} = \vec{A}\vec{x} + \vec{b},
\end{aligned}$
}
\end{align}
where $\vec{x}\in\Complex^N$ is an $N$-pixel image that has been vectorized, $\vec{y}\in\Complex^M$ is the MRI data measured from $C$ receive coils, $\vec{b}\in\Complex^M$ is circularly symmetric white Gaussian noise with variance $\sigma^2$, and $\vec{A}\in \Complex^{M\times N}$ is a known forward operator that incorporates pixel-wise multiplication with coil sensitivity maps, discrete Fourier transform, and k-space undersampling. 

When applying DIP method to MRI reconstruction, a random code vector $\vec{z}$ is fed into a network $\mathcal{\vec{G}}_{\vec{\theta}}(\cdot)$ to produce an estimate of the true image $\vec{x}$. The network parameters $\vec{\theta}$ and the code vector $\vec{z}$ are then optimized to make the network output consistent with the measured k-space data $\vec{y}$ using a known forward operator $\vec{A}$. This optimization process is formulated as

\begin{align}
\scalebox{0.8}{%
$\begin{aligned}
\hvec{z}, \hvec{\theta} = \argmin_{\vec{z}, \vec{\theta}} {\left\| \vec{A}\mathcal{\vec{G}}_{\vec{\theta}}(\vec{z})-\vec{y} \right\|_2^2}.
\end{aligned}$
}
\label{eq:dip}
\end{align}

After training the network, the final image is reconstructed using $\hvec{x}=\mathcal{\vec{G}}_{\hvec{\theta}}(\hvec{z})$. It is important to note that DIP does not rely on an explicit prior in the form of a regularization term but instead exploits the inherent structure of the network to produce images that appear natural.

\subsection{\discus}
\begin{figure}[!h]
  \centering
  \begin{tikzpicture}
    \node[anchor=north west, inner sep=0] (image) at (0,-0.25) {\includegraphics[width=0.47\textwidth]
    {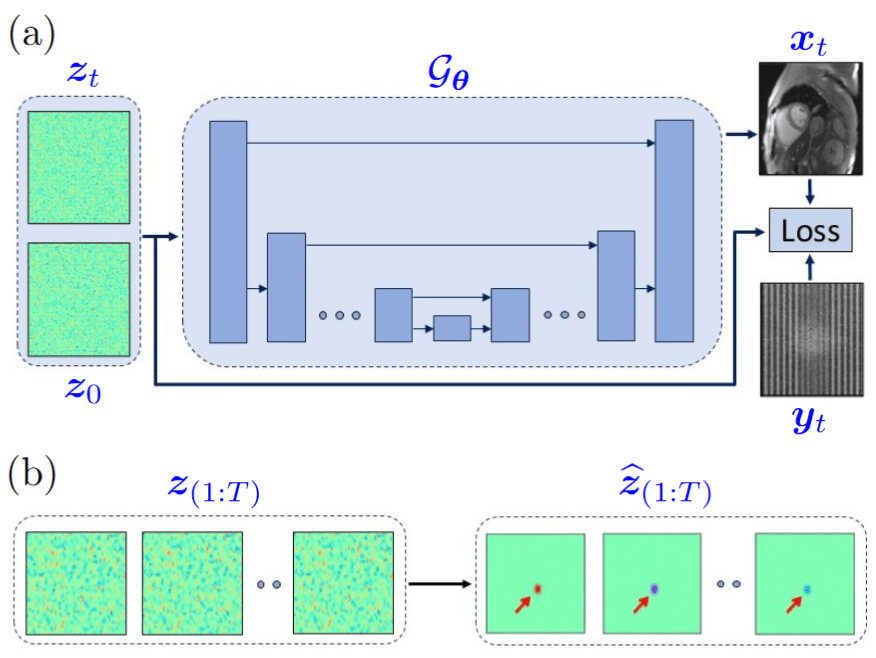}};
  \end{tikzpicture}
  \caption{\textr{(a) Overview of the \discus framework: A U-Net $\mathcal{\vec{G}}_{\vec{\theta}}$ is fed one static $\vec{z}_0$ and a dynamic $\vec{z}_{t}$ code vector to generate an estimate of the frame $\vec{x}_{t}$, which is made consistent with the measured data $\vec{y}_{t}$. The training is sequentially repeated for all frames in the image series. (b) An example of dynamic code vectors $\vec{z}_{(1:T)}$ before (left) and after (right) training.} \textb{The support of the dynamic code vectors, i.e., the number of non-zero entries in the temporal union of $\vec{z}_{(1:T)}$ (one in this case), defines the dimensionality of the manifold. These non-zero entries generate the temporal variation in the image series.}}
  \label{fig:discus-framework}
\end{figure}

 The high-level description of \discus is provided in \figref{discus-framework}. \discus attempts to construct an image series with $T$ total frames using a single network and time-varying code vectors. It is inspired by DIP but enforces additional constraints on the code vectors. The image series in \discus is recovered by solving the following optimization problem:

\begin{align}
\scalebox{0.8}{%
$\begin{aligned}
\hvec{z}_0, \hvec{z}_{(1:T)}, \hvec{\theta} = \argmin_{\vec{z}_0, \vec{z}_{(1:T)}, \vec{\theta}} & \sum_{t=1}^{T} \left\| \vec{A}_t \mathcal{\vec{G}}_{\vec{\theta}}(\vec{z}_0, \vec{z}_t) - \vec{y}_t \right\|_2^2 \\
& \hspace{1.5cm} + \lambda \left\| \vec{z}_{(1:T)} \right\|_{2,1},
\end{aligned}$%
}
\label{eq:discus}
\end{align}

where $\vec{y}_t\in\Complex^M$, $\vec{x}_t\in\mathbb{C}^N$, and $\vec{A}_t\in\mathbb{C}^{M\times N}$ are measured k-space data, image frame, and forward operator for the $t^{\sf th}$ frame, respectively. 
\textr{In addition, $\vec{z}_0\in\Real^{k \times N}$ represents static code vectors with $k$ channels, while $\vec{z}_t\in\Real^{N}$ represents single-channel dynamic code vectors at time $t$.} 
We denote the temporal sequence of an arbitrary variable $(\cdot)$ with $T$ entries as $(\cdot)_{(1:T)}$. Finally,  $\left\| \vec{z}_{(1:T)}\right\|_{2,1} = {\sum_{n=1}^N\sqrt{ \scriptstyle \sum_{t=1}^T z_t^2[n]}}$ is a hybrid $\ell_2$-$\ell_1$ norm that first computes elementwise $\ell_2$ norm along the time dimension followed by the $\ell_1$ norm along the remaining dimensions. Note that $z_t[n]$ represents the $n{\text{th}}$ element in $\vec{z}_t$, and $\lambda>0$ is a constant that controls the strength of group sparsity.

\textr{In \discus, both $\vec{z}_0$ and $\vec{z}_t$ are learnable parameters. Before being fed into the generator $\mathcal{\vec{G}}_{\vec{\theta}}(\cdot)$, they are concatenated along the channel dimension, forming $(k+1)$-channel network input $\vec{z} \in \mathbb{R}^{(k+1) \times N}$ such that $\mathcal{\vec{G}}_{\vec{\theta}} \colon \mathbb{R}^{(k+1) \times N}\rightarrow \mathbb{R}^{2 \times N}$ at any time $t$.}
As shown in \figref{discus-framework}(a), the generator maps the code vectors to an estimate of $\vec{x}_t$. The network parameters and code vectors are jointly optimized using the objective function in \eqref{discus}. The static code vector $\vec{z}_0$ is optional but enhances the expressivity of the network by providing additional degrees of freedom. The dynamic code vectors $\vec{z}_{(1:T)}$ capture the frame-to-frame variations in the image series. The second term in \eqref{discus} imposes a hybrid $\ell_2$-$\ell_1$ norm that enforces group sparsity \cite{yuan2006groupsparsity} on $\vec{z}_{(1:T)}$. Group sparsity ensures that the dynamic code vectors are not only sparse but also share a common support across all time frames, aiding in manifold discovery.
\textr{Since the intrinsic manifold dimensionality is expected to be much smaller than $N$, we set $k=1$ for $\vec{z}_t$. Additionally, our method was found to be insensitive to small variations in $k$ for $\vec{z}_0$, with values in the range $2 < k \leq 8$ yielding similar performance. Based on these empirical observations, we choose $k=3$ for $\vec{z}_0$.} 
After training, the number of non-zero entries in each $\hvec{z}_t$ determines the dimensionality of the underlying manifold, while the values of these non-zero entries capture temporal variations over time. A visual representation of a one-dimensional manifold is shown in \figref{discus-framework}(b), where all entries in $\hvec{z}_t$ are zero except for one. Notably, the manifold dimension is not predefined; instead, \discus dynamically discovers it during training based on the given data. Once the network is trained, the reconstructed image for the $t^\text{th}$ frame is obtained as $\hvec{x}_t = \mathcal{\vec{G}}_{\hvec{\theta}} (\hvec{z}_0, \hvec{z}_t)$.

Although \discus can be applied to various applications, this work focuses on free-breathing single-shot LGE imaging. In this application, the same imaging plane is acquired once every heartbeat or every other heartbeat. The long temporal delays between consecutive frames cause large variations in the respiratory phase, which is assumed to be unknown. A common strategy involves reconstructing each frame separately and then averaging them after registration to enhance the signal-to-noise ratio (SNR) \cite{piehler2013free}. In contrast, manifold learning offers a unified framework to recover all frames jointly \cite{poddar2015SToRM}. Under ideal conditions, a manifold of dimensionality one is optimal for this application. However, residual cardiac motion caused by heart rate variability, intensity changes due to variations in RF timing, and/or peristalsis are invariably present in LGE imaging. These factors make the true manifold dimensionality subject-specific and inherently variable. 
Compared to previous work on manifold learning, the group sparsity in \discus provides a data-driven approach for selecting the manifold dimensionality.

\subsection{Study I--Dynamic Phantom}
\textr{In this study, we used a $128\times 128$ complex-valued Shepp-Logan phantom to demonstrate that \discus can discover the dimensionality of the underlying manifold by imposing group sparsity on dynamic code vectors.} We generated three distinct image series, each with $T=64$ frames. The first series originated from a manifold of dimensionality one and involved random rotations within $\pm 3^\circ$ relative to the first frame. The second series also originated from a manifold of dimensionality one and involved random horizontal translations within $\pm3$ pixels relative to the first frame. The final series originated from a manifold of dimensionality two, with each frame perturbed both by random rotations and random translations. To simulate single-coil k-space data, each frame was Fourier transformed. Then, white Gaussian noise was added to the k-space data to yield an SNR of 25 dB. The simulated noisy k-space data were subsequently undersampled with a Cartesian mask. The 12 central phase-encoding (PE) indices were fully sampled, while the outer 52 PE indices were sampled uniformly at random at an acceleration rate of $R=2$. The frequency encoding (FE) dimension was fully sampled. The reference image series was created using the fully sampled noiseless data, and it was then used to quantitatively evaluate DISCUS in terms of normalized mean squared error (NMSE), expressed in dB, and structural similarity index measure (SSIM). 
To further assess the consistency of the discovered manifold dimensionality, we repeated the \discus reconstruction ten times, each time with a different initialization of network weights $\vec{\theta}$ and the code vectors $\vec{z}_0$ and $\vec{z}_t$.

\subsection{Study II--LGE Simulation}
In this study, we simulated six different free-breathing LGE image series from three distinct male and three distinct female MRXCAT subjects \cite{wissmann2014mrxcat}.  
\textr{To incorporate breathing-induced motion, a unique breathing pattern was simulated for each image series by parametrically modeling the diaphragm hysteresis curve \cite{lapshin1995hysterisis} and randomly selecting the starting point along the curve.} 
To enhance realism, myocardial scars were added to three of the six series. 
\textb{The parameters defining tissue contrast, including proton density (PD), T1, inversion time (TI), and gadolinium (Gd) concentration, were selected to match routine LGE scans at our institution. Specifically, we used PD = 80\%, T1/T2 = 955/33 ms for myocardium and 1411.1/248.8 ms for blood, TR = 3 ms, and TI $\approx$ 711.9 ms (computed as $1.05 \times \ln(2) \times \text{T}1_{\text{myocardium}}$).}
A $224 \times 192$ short-axis slice was selected from each digital subject with $T=32$ frames. We simulated $8$ complex-valued coil sensitivity maps using the Biot-Savart law. White Gaussian noise was added to the multi-coil k-space data to yield an SNR of 25 dB. 
\textr{The resulting noisy data were retrospectively undersampled at $R = 2,~3,~4,~5,~\text{and}~6$ using the golden ratio offset (GRO) Cartesian sampling mask with fully sampled readout \cite{joshi2022GRO}}. 
The coil-combined reference from fully sampled data was used for quantitative assessment of \discus in terms of NMSE and SSIM. 

For the second part of study II, we performed an ablation study to assess the impact of the number of frames and group sparsity on the image quality. To this end, we performed the \discus reconstruction with 8, 16, and 32 frames, naming the resulting reconstructions as \discus-8, \discus-16, and \discus-32, respectively. For comparison, low-rank + sparse (L+S) reconstructions \cite{otazo2015LplusS} were also performed with 8, 16, and 32 frames, labeled as (L+S)-8, (L+S)-16, and (L+S)-32, respectively. To study the contribution of group sparsity, we implement \discus without group sparsity (DGS) and repeated the DGS reconstruction with 8, 16, and 32 frames, resulting in DGS-8, DGS-16, and DGS-32, respectively.

\subsection{Study III--LGE with Retrospective Undersampling}
Eight single-shot free-breathing LGE series, each with $T=32$ frames, were collected from clinical patients on a commercial 1.5T scanner (MAGNETOM Sola, Siemens Healthcare, Erlangen, Germany) using phase-sensitive inversion recovery (PSIR) sequence yielding two sets of measurements: T1-weighted (T1) and proton density (PD) with balanced steady-state free precession (bSSFP) readout \cite{kellman2002PSIR}. The multi-coil data had $16$ to $24$ coils with matrix sizes of $160 \times 96$ to $160 \times 116$ and were collected in short-axis (SAX) and three-chamber (3CH) views. The imaging parameters were set as this: spatial resolution $(2.19-2.5) \times (2.74-3.13)$ mm$^2$, slice thickness $8$ mm, temporal footprint $249.6$ to $299$ ms, echo time $1.26$ to $1.3$ ms, inversion time $330$ to $430$ ms, and flip angle $40^\circ$. The acceleration rate of $R=1$ was enabled by lower spatial resolutions and longer temporal footprints. Although not realistic, this approach allowed us to compute NMSE and SSIM using the fully sampled images as a reference. 
\textr{Readout oversampling in the raw data was removed, and noise pre-whitening was performed using noise scan. The data were compressed to eight virtual coils using principal component analysis \cite{buehrer2007PCA}, and coil sensitivities were estimated using ESPIRiT \cite{uecker2014espirit}.}
To suppress the undesired brightness variations, surface coil correction was performed by adjusting the coil sensitivity maps before reconstruction \cite{lei2023SCC}. 
\textr{Each image series was retrospectively undersampled using the GRO sampling mask at $R=2,~3,~4,~5,$ and $6$ \cite{joshi2022GRO}}. 
The coil-combined reference from the fully sampled data was used for the quantitative assessment of different methods.

\vspace{0.5cm} 

\subsection{Study IV--LGE with Prospective Underampling}
Eight single-shot free-breathing image series, each with $T=32$ frames, were collected from clinical patients on a commercial 1.5T scanner (MAGNETOM Sola, Siemens Healthcare, Erlangen, Germany) using a phase-sensitive inversion recovery LGE sequence \cite{kellman2002PSIR}. The data were prospectively undersampled at $R=4$ to $6$ using GRO pattern \cite{joshi2022GRO} and had matrix sizes of $256 \times 192$ to $256 \times 288$. The imaging parameters were: spatial resolution $(1.22-1.37) \times 1.37$ mm$^2$, slice thickness 8 mm, temporal footprint $124.8$ ms, echo time $1.2$ ms, inversion time $330$ to $410$ ms, and flip angle $40^\circ$. 
\textr{To mimic the common clinical practice, the individual LGE images were motion corrected (MoCo) \textb{\cite{xue2012moco}} independently for each method and then time-averaged.}

For both LGE studies III and IV, the PD and T1-weighted measurements were coil-compressed jointly; however, for all DIP-based methods, PD images were reconstructed using CS for faster processing. Note, these PD images are used to phase-correct T1-weighted images and are expected to be smooth. We did not observe any benefit of reconstructing PD images using DIP-based methods.

\subsection{Implementation Details}
In DISCUS and other competing DIP-based methods, we used a U-Net with 6 layers and 128 channels per layer, following the architecture from the original DIP implementation \cite{ulyanov2018DIP}, which includes noise regularization.
\textr{In our configuration, we used 6 downsampling blocks that performed strided convolutions and 6 upsampling blocks that utilized nearest neighbor interpolation. Each block included batch normalization layers and leaky ReLU as the activation function.}
\textr{This architecture contained 3,591,682 trainable parameters.}
\textr{We chose a 2-channel real-valued processing to handle complex data.}
\textr{We normalized undersampled k-space data to a scale between 0 and 10 before training.}
\textr{For DISCUS, we trained a separate network for each image series within a study.}
\textr{We used Adam optimizer with batch size $8$ and step scheduling for learning rate with starting learning rate $0.001$, step size $500$ iterations, and decaying gamma rate $0.97$.}
The network in \discus was trained for 10,000 iterations, requiring 160, 90, and 75 minutes on a single NVIDIA RTX3090 GPU to reconstruct an entire image series in Studies I, II, and III/IV, respectively. In comparison, DIP required 150 and 123 minutes for Studies II and III/IV, and self-guided DIP (SG-DIP) required 102 and 74 minutes for Studies II and III/IV on the same hardware. All four studies can be reproduced using the code and data available at \url{https://github.com/OSU-MR/discus}. 

\subsection{Evaluation}
In all four numerical studies, we compared \discus with CS \cite{lustig2007CS}, L+S \cite{otazo2015LplusS}, the original DIP \cite{ulyanov2018DIP}, and the recently proposed SG-DIP \cite{bell2023SG_DIP}. For CS, we imposed $\ell_1$-norm minimization in the spatial wavelet domain. The L+S algorithm used temporal FFT as a sparsifying transform. In the two competing DIP-based methods i.e., DIP and SG-DIP, we reconstructed each frame individually by training $T$ separate networks in each image series. 
The free parameters across all methods, including $\lambda$, learning rate, and number of iterations, were optimized based on NMSE using an additional fully sampled dataset from the retrospective study. 

\textb{In studies I, II, and III, where we had a fully sampled reference, we evaluated the reconstruction performance on coil-combined images using NMSE and SSIM. NMSE was defined as $20\log_{10}\left(\|\vec{x}_t - \hvec{x}_t\|_2 / \|\vec{x}_t\|_2\right)$ for the $t^{\sf th}$ complex-valued frame and was then averaged over all $T$ frames. Likewise, SSIM \cite{wang2004SSIM} was first computed individually for each magnitude-only frame using the function provided by the scikit-image Python package, and then averaged over all $T$ frames.} 
For Study IV, where the fully sampled reference was not available, MoCo was applied to register all reconstructed frames to a single image. The resulting motion-corrected images were blindly scored by two expert cardiac MRI readers, each with more than 10 years of experience in cardiac MRI. Each LGE image was scored on a five-point Likert scale (1: Non-diagnostic, 2: Poor, 3: Adequate, 4: Good, and 5: Excellent). For each patient, the readers were also instructed to select the best image or images in terms of overall image quality. This was done to ensure a more granular evaluation of the image quality in cases where multiple reconstructions from the same patient received the highest score.

\section{Results}\label{sec:res}
\subsection{Study I--Dynamic Phantom}
\begin{figure*}[!h]
  \centering
  \begin{tikzpicture}
    \node[anchor=south west, inner sep=0] (image) at (-0.2,1.45) {\includegraphics[width=0.95\textwidth]{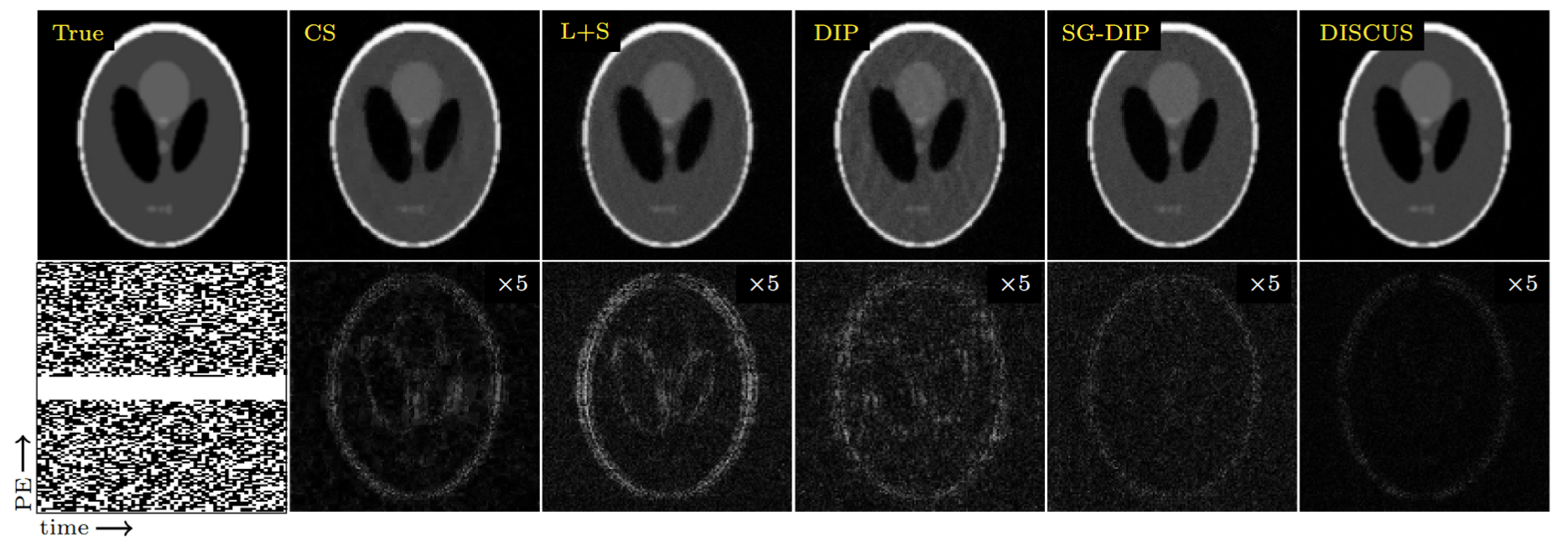}};
  \end{tikzpicture}
\caption{Representative results from the Shepp-Logan phantom (Study I). First row shows an example frame with reference image (left) and reconstructions by CS, L+S, DIP, SG-DIP, and \discus, from the series containing both rotations and translations. The second row contains the sampling pattern (left) where frequency-encoding is not displayed, and $\times 5$ error maps associated with reconstructions in the first row.}
  \label{fig:Shepp_Logan_fig}
\end{figure*}

In this study, we repeated \discus reconstruction $10$ times for three different image series, i.e., only rotations, only translation, and both random rotations and translations. \textr{We report the mean and standard error of the mean (SEM) over the 10 repetitions, where SEM is defined as SD$/\sqrt{N}$, with SD representing the standard deviation and $N$ the number of repetitions.}
The NMSE and SSIM values are summarized in \tabref{Shepp_Logan_tab}, with \discus outperforming other methods by a wide margin for all three image series.
\textb{For the first two image series with only rotations and only translations, \discus successfully identified a one-dimensional manifold in all trial runs.}
However, for the last image series with rotations and translations, \discus discovered the true manifold dimensionality of two in eight out of the ten repetitions. 
 
In the other two instances, the discovered manifold dimensionality was three, one more than the true value. We attribute this to the highly nonconvex cost surface occasionally leading to imperfect disentanglement of the manifold dimensions. A representative frame from a series with both rotations and translations is shown in \figref{Shepp_Logan_fig}. In this example, CS exhibits blocky artifacts around the edges, while L+S and DIP show noise amplification and residual motion artifacts. Being less prone to overfitting, SG-DIP outperforms DIP in terms of noise amplification. However, \discus outperforms all methods in terms of noise amplification and preserving sharp edges. The movies corresponding to \figref{Shepp_Logan_fig}, showing both dynamic code vectors $\hvec{z}_t$ and image series $\hvec{x}_t$, are shown in Online Resource 1.

\begin{table}[!h]
\centering
\renewcommand{\arraystretch}{1.5}
\begin{adjustbox}{width=0.48\textwidth}
\begin{tabular}{|ll|ccccc|}
\hline
& & \textbf{CS} &\textbf{L+S} &\textbf{DIP} &\textbf{SG-DIP} &\textbf{\discus} \\
\hline
\multirow{2}{*}{\rotatebox{90}{\scriptsize Rot. }}
& SSIM & 0.883 & 0.831 & 0.806 & 0.853 & \textbf{0.961} \textr{$\pm$ 0.0005} \\
& NMSE & -23.82 & -20.30 & -21.03 & -26.57 & \textbf{-31.02} \textr{$\pm$ 0.045} \\
\hline
\multirow{2}{*}{\rotatebox{90}{\scriptsize Tran. }}
& SSIM & 0.883 & 0.835 & 0.799 & 0.854 & \textbf{0.960} \textr{$\pm$ 0.0005} \\
& NMSE & -23.75 & -20.54 & -20.69 & -25.31 & \textbf{-30.70} \textr{$\pm$ 0.054} \\
\hline
\multirow{2}{*}{\rotatebox{90}{\scriptsize Both }}
& SSIM & 0.882 & 0.834 & 0.798 & 0.846 & \textbf{0.923} \textr{$\pm$ 0.0004} \\
& NMSE & -23.73 & -19.20 & -20.47 & -24.65 & \textbf{-28.66} \textr{$\pm$ 0.01} \\
\hline
\end{tabular}
\end{adjustbox}
\caption{\textr{Quantitative results from the Shepp-Logan phantom (Study I). The rows show NMSE and SSIM values for the image series with rotations (top), translations (center), and a combination of both (bottom). The last column contains average ($\pm$SEM) values from 10 repetitions of \discus. Bold values indicate the best results.}}
\label{tab:Shepp_Logan_tab}
\end{table}

\subsection{Study II--LGE Simulation}

\begin{figure*}[!h]
  \centering
  \begin{tikzpicture}[scale=1]
    \node[anchor=north west, inner sep=0] (image) at (0,0) {\includegraphics[width=0.65\textwidth]{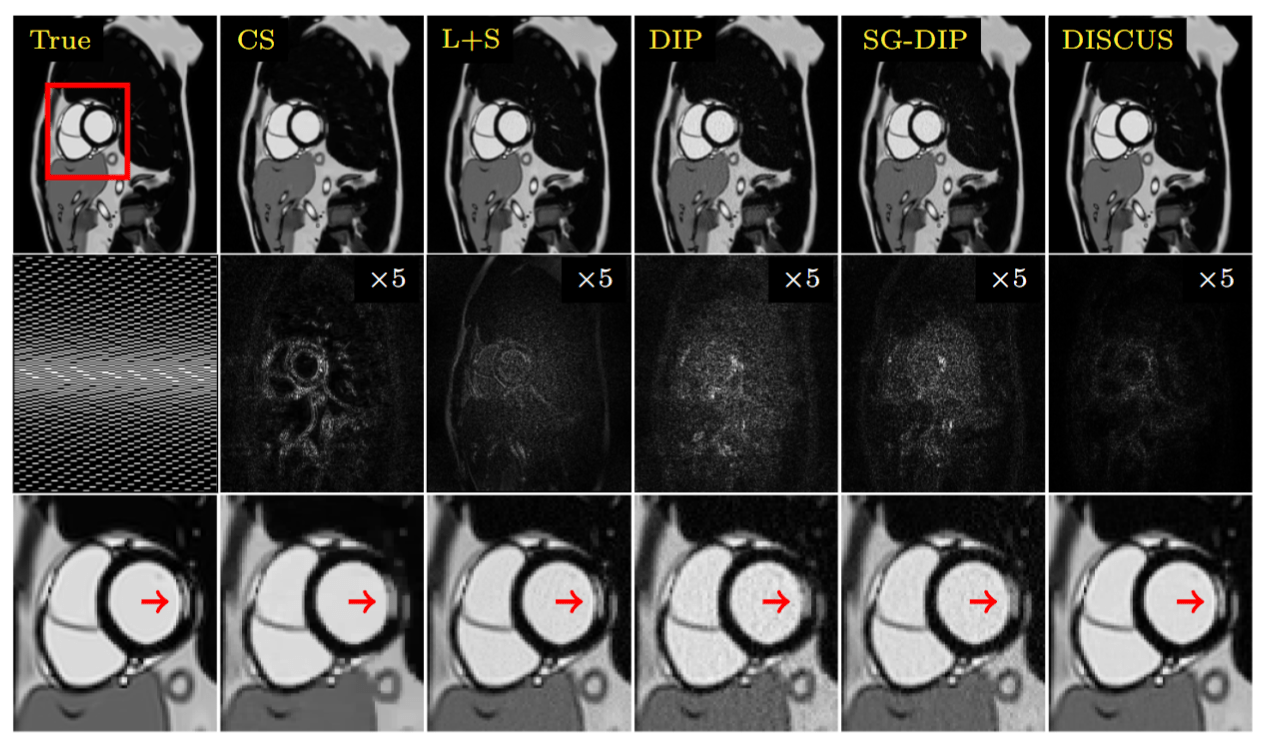}};
    
  \end{tikzpicture}
  \caption{Representative results from the simulated LGE (Study II) exhibiting a pronounced myocardial scar at $R=4$. First row shows an example frame with reference (left) and reconstructions by CS, L+S, DIP, SG-DIP and \discus. Second row contains the GRO sampling pattern (left) where frames are displayed left-to-right, phase encoding is shown top-to-bottom, and frequency encoding is omitted, and $\times 5$ error maps. The final row provides a zoomed-in view of the red box in first row, with the red arrows pointing to the scar.}
  \label{fig:MRXCAT_comparison_fig}
\end{figure*}

Using six simulated LGE images, we compared CS, L+S, DIP, SG-DIP, and \discus at four different acceleration rates, i.e., $R=2,~3,~4,~\text{and}~5$. The results are summarized in \tabref{MRXCAT_comparison_tab}. \discus outperforms other methods by a significant margin. In particular, at the highest acceleration rate of $R=5$, the NMSE advantage of \discus over the second best method (CS) is greater than 5 dB.  \figref{MRXCAT_comparison_fig} presents a representative frame from one of the image series with a simulated scar. In this $R=4$ example, CS excessively smoothens the scar, while DIP and SG-DIP exhibit noise amplification. Both L+S and \discus preserve the scar's conspicuity, with L+S showing slightly more noise amplification inside the blood pool. The error images also reveal that L+S loses some edge information, even though this is not apparent in the reconstructed frame. 

\begin{figure*}[!h]
  \centering
  \begin{tikzpicture}
    \node[anchor=north west, inner sep=0] (image) at (-0.2,0) {\includegraphics[width=0.99\textwidth]{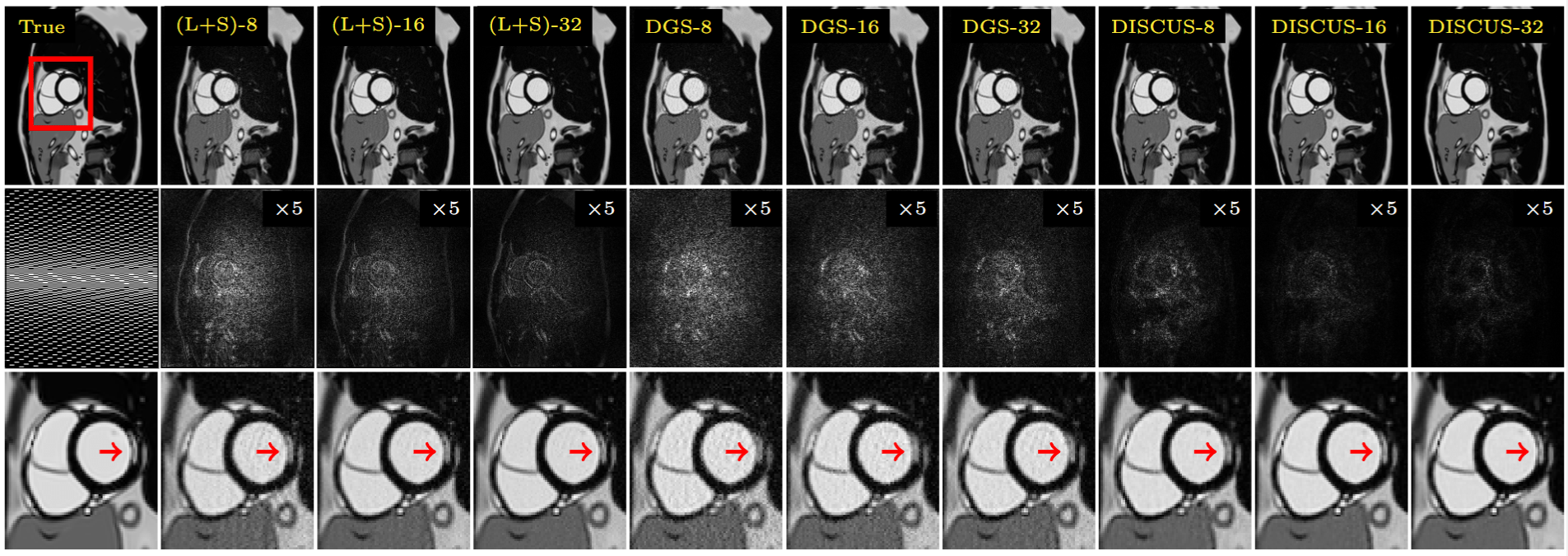}};

  \end{tikzpicture}
  \caption{Representative results from the ablation study of the simulated LGE (Study II) exhibiting a pronounced myocardial scar at $R=4$. First row shows an example frame with reference (left) and reconstructions by L+S, DGS, and \discus with 8, 16, and 32 frames. Second row contains the GRO sampling pattern (left) where frames are displayed left-to-right, phase encoding is shown top-to-bottom, and frequency encoding is omitted, and $\times 5$ error maps. The final row provides a zoomed-in view of the red box in first row, with the red arrows pointing to the scar.}
  \label{fig:MRXCAT_ablations_fig}
\end{figure*}

The findings of the ablation study, where we investigate the impact of $T$ and the group sparsity, are summarized in \tabref{MRXCAT_ablations_tab}. In this study, we only compare methods that jointly recover all the frames, i.e., L+S, \discus, and \discus without group sparsity (DGS). As expected, the performances of all three methods degrade with a decrease in $T$, indicating that these methods exploit the joint information across the frames. Also, across all three values of $T$, \discus maintains an advantage of more than 3 dB over DGS. Since DGS is identical to \discus except for the group sparsity, these numbers highlight the impact of manifold learning in \discus. The dimensionality of the manifold discovered by \discus was either one or two across all six simulated LGE images. Forcing the manifold dimensionailty to one resulted in degradation of reconstruction quality.
A representative image from the ablation study is shown in \figref{MRXCAT_ablations_fig}. Both DGS and L+S exhibit noise amplification, especially at $T=8$. 


\begin{table*}[!h]
\centering
\renewcommand{\arraystretch}{1.5}
\begin{adjustbox}{width=0.8\textwidth}
\begin{tabular}{|ll|ccccc|}
\hline
& & \textbf{CS} &\textbf{L+S} &\textbf{DIP} &\textbf{SG-DIP} &\textbf{\discus} \\
\hline
\multirow{2}{*}{\rotatebox{90}{\scriptsize $R=2$ }}
& SSIM & 0.931 \textr{$\pm$ 0.0059} & 0.932 \textr{$\pm$ 0.0031} & 0.932 \textr{$\pm$ 0.0029} & 0.960 \textr{$\pm$ 0.0021} & \textbf{0.980} \textr{$\pm$ 0.0018}  \\
& NMSE & -27.49 \textr{$\pm$ 0.231} & -24.63 \textr{$\pm$ 0.432} & -26.15 \textr{$\pm$ 0.356} & -27.60 \textr{$\pm$ 0.265} & \textbf{-29.64} \textr{$\pm$ 0.249}\\
\hline
\multirow{2}{*}{\rotatebox{90}{\scriptsize $R=3$ }}
& SSIM & 0.921 \textr{$\pm$ 0.0067} & 0.928 \textr{$\pm$ 0.0030} & 0.889 \textr{$\pm$ 0.0047} & 0.932 \textr{$\pm$ 0.0024} & \textbf{0.979} \textr{$\pm$ 0.0019} \\
& NMSE & -26.14 \textr{$\pm$ 0.313} & -23.26 \textr{$\pm$ 0.552} & -23.78 \textr{$\pm$ 0.467} & -24.81 \textr{$\pm$ 0.313} & \textbf{-28.96} \textr{$\pm$ 0.313}\\
\hline
\multirow{2}{*}{\rotatebox{90}{\scriptsize $R=4$ }}
& SSIM & 0.905 \textr{$\pm$ 0.0077} & 0.922 \textr{$\pm$ 0.0034} & 0.846 \textr{$\pm$ 0.0053} & 0.903 \textr{$\pm$ 0.0035} & \textbf{0.978} \textr{$\pm$ 0.0034} \\
& NMSE & -24.25 \textr{$\pm$ 0.330} & -22.44 \textr{$\pm$ 0.776} & -21.66 \textr{$\pm$ 0.590} & -22.74 \textr{$\pm$ 0.380} & \textbf{-28.03} \textr{$\pm$ 0.318}\\
\hline
\multirow{2}{*}{\rotatebox{90}{\scriptsize $R=5$ }}
& SSIM & 0.886 \textr{$\pm$ 0.0088} & 0.903 \textr{$\pm$ 0.0047} & 0.816 \textr{$\pm$ 0.0057} & 0.858 \textr{$\pm$ 0.0041} & \textbf{0.969} \textr{$\pm$ 0.0042} \\
& NMSE & -22.19 \textr{$\pm$ 0.409} & -20.28 \textr{$\pm$ 0.833} & -19.01 \textr{$\pm$ 0.602} & -20.85 \textr{$\pm$ 0.399} & \textbf{-27.96} \textr{$\pm$ 0.436}\\ 
\hline
\multirow{2}{*}{\rotatebox{90}{\scriptsize \textr{$R=6$}}} 
& \textr{SSIM} & \textr{0.868} \textr{$\pm$ 0.0107} & \textr{0.902} \textr{$\pm$ 0.0060} & \textr{0.758} \textr{$\pm$ 0.0077} & \textr{0.813} \textr{$\pm$ 0.0053} & \textbf{\textr{0.958}} \textr{$\pm$ 0.0045}  \\
& \textr{NMSE} & \textr{-20.25} \textr{$\pm$ 0.489} & \textr{-19.44} \textr{$\pm$ 0.854} & \textr{-16.58} \textr{$\pm$ 0.635} & \textr{-17.7} \textr{$\pm$ 0.453} & \textbf{\textr{-27.59}} \textr{$\pm$ 0.476}\\
\hline
\end{tabular}
\end{adjustbox}
\caption{\textr{Quantitative results from the simulated LGE (Study II). The rows show NMSE and SSIM values at $R=2, 3, 4, 5,$ and $6$. Average values ($\pm$SEM) are presented from six distinct image series. Bold values indicate the best results.}}
\label{tab:MRXCAT_comparison_tab}
\end{table*}


\begin{table}[!h]
\centering
\renewcommand{\arraystretch}{1.75}
\begin{adjustbox}{width=0.47\textwidth}
\begin{tabular}{|ll|ccc|}
\hline
& & \textbf{L+S}  & \textbf{DGS} & \textbf{\discus} \\
\hline
\multirow{2}{*}{\rotatebox{90}{\scriptsize $T=8$ }}
& SSIM   & 0.827 \textr{$\pm$ 0.0065}  & 0.849 \textr{$\pm$ 0.0091} & \textbf{0.952} \textr{$\pm$ {0.0064}} \\
& NMSE   & -19.82 \textr{$\pm$ 0.679} & -20.18 \textr{$\pm$ 0.348} & \textbf{-24.30} \textr{$\pm$ {0.343}} \\
\hline
\multirow{2}{*}{\rotatebox{90}{\scriptsize $T=16$}}
& SSIM & 0.879 \textr{$\pm$ 0.0047} & 0.875 \textr{$\pm$ 0.0063} & \textbf{0.969} \textr{$\pm$ {0.0046}} \\ 
& NMSE & -21.22 \textr{$\pm$ 0.718} & -22.42 \textr{$\pm$ 0.314} & \textbf{-26.06} \textr{$\pm$ {0.304}} \\
\hline
\multirow{2}{*}{\rotatebox{90}{\scriptsize $T=32$}}
& SSIM & 0.922 \textr{$\pm$ 0.0034} & 0.902 \textr{$\pm$ 0.0049} & \textbf{0.978} \textr{$\pm$ {0.0034}} \\ 
& NMSE & -22.44 \textr{$\pm$ 0.776} & -23.43 \textr{$\pm$ 0.373} & \textbf{-28.03} \textr{$\pm$ {0.318}} \\
\hline

\end{tabular}
\end{adjustbox}
\caption{\textr{Quantitative results from the ablations performed on simulated LGE (Study II) at $R=4$. The rows show NMSE and SSIM values at $T=8, 16,$ and $32$. Average values ($\pm$SEM) are presented from six distinct image series. Bold values indicate the best results.}}
\label{tab:MRXCAT_ablations_tab}
\end{table}

\subsection{Study III--LGE with Retrospective Undersampling}

\begin{figure*}[!h]
  \centering
  \begin{tikzpicture}[scale=1]
    \node[anchor=north west, inner sep=0] (image) at (0,0) 
    {\includegraphics[width=0.81\textwidth]{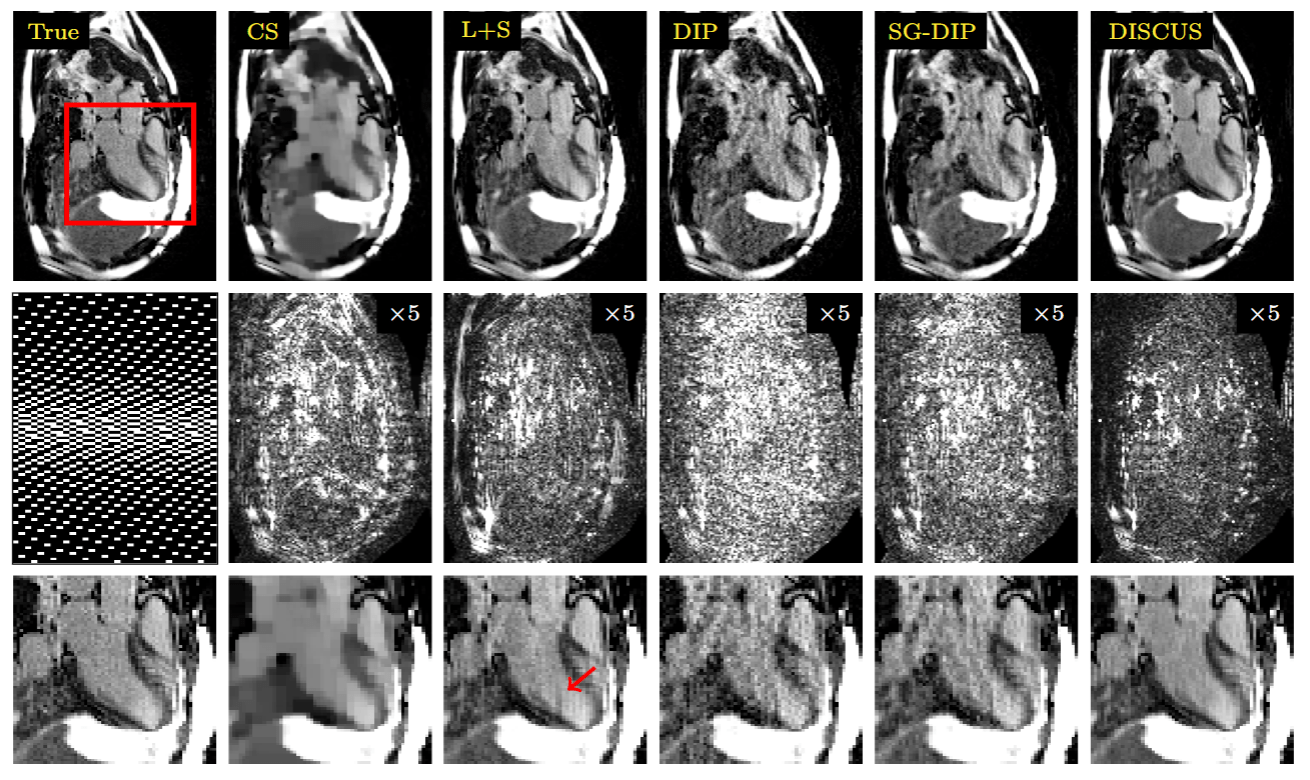}};
  \end{tikzpicture}
  \caption{Representative results from the retrospective patient LGE (Study III) at $R=4$. First row shows an example frame from one of the 8 patients with reference (left) and reconstructions by CS, L+S, DIP, SG-DIP and \discus. Second row contains the GRO sampling pattern (left) where frames are displayed left-to-right, phase encoding is shown top-to-bottom, and frequency encoding is not displayed, and $\times 5$ error maps. The final row provides a zoomed-in view of the red box in first row, with the artifact in L+S reconstruction highlighted by red arrow.}
  \label{fig:LGE_retro_fig}
\end{figure*}

\begin{figure*}[!h]
  \centering
  \begin{tikzpicture}[scale=1]
    \node[anchor=north west, inner sep=0] (image) at (0,0) 
    {\includegraphics[width=0.81\textwidth]{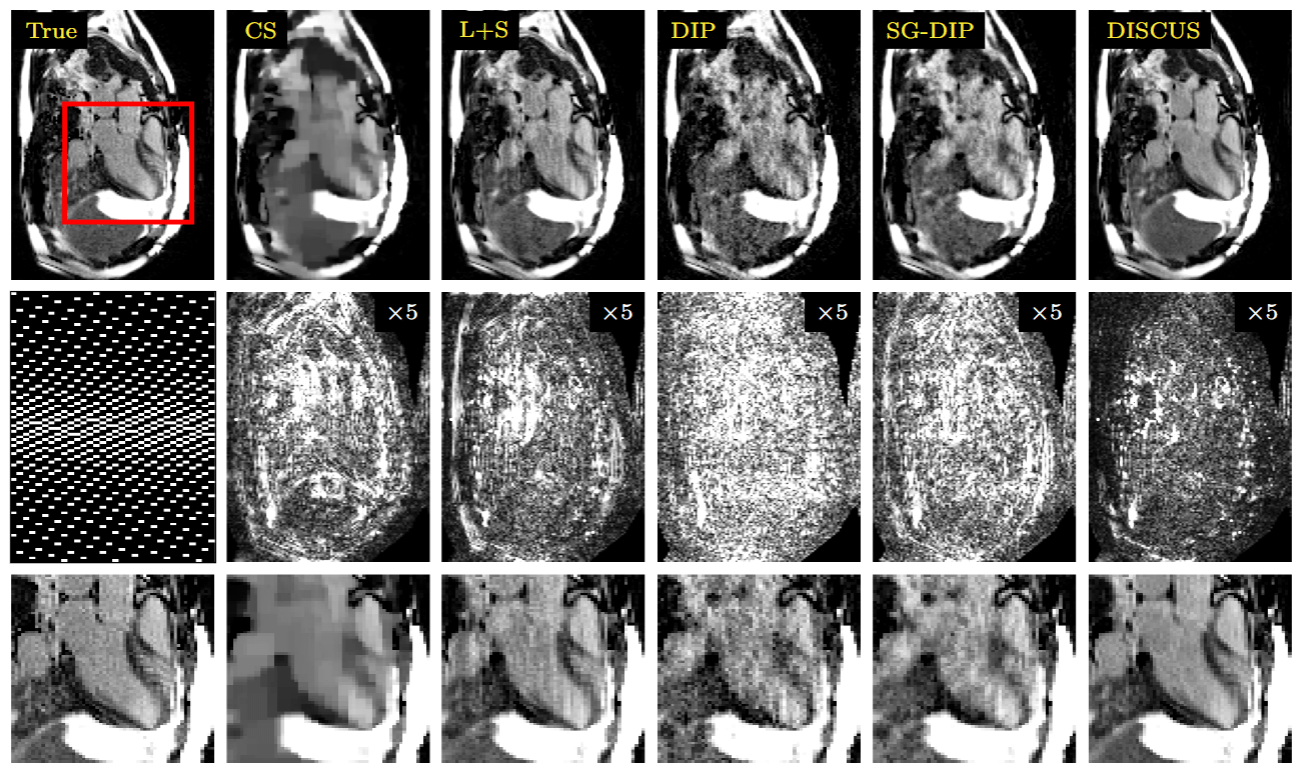}};

  \end{tikzpicture}
  \caption{Representative results from the retrospective patient LGE (Study III) from the same patient as in \figref{LGE_retro_fig} but at a higher acceleration rate, i.e., $R=5$. First row shows an example frame with reference (left) and reconstructions by CS, L+S, DIP, SG-DIP, and \discus. Second row contains the GRO sampling pattern (left) where frames are displayed left-to-right, phase encoding is shown top-to-bottom, and frequency encoding is not displayed, and $\times 5$ error maps. The final row provides a zoomed-in view of the red box in first row. CS and DIP reconstructions completely break down here.}
  \label{fig:LGE_retro_R5_fig}
\end{figure*}

\begin{table*}[!h]
\centering
\renewcommand{\arraystretch}{1.5}
\begin{adjustbox}{width=0.8\textwidth}
\begin{tabular}{|ll|ccccc|}
\hline
& & \textbf{CS} &\textbf{L+S} &\textbf{DIP} &\textbf{SG-DIP} &\textbf{\discus} \\
\hline
\multirow{2}{*}{\rotatebox{90}{\scriptsize $R=2$}}
& SSIM & 0.956 \textr{$\pm$ 0.0054}  & \textbf{0.979} \textr{$\pm$ 0.0017} & 0.950 \textr{$\pm$ 0.0023} & 0.965 \textr{$\pm$ 0.0029} & 0.977 \textr{$\pm$ 0.0016}  \\
& NMSE & -21.84 \textr{$\pm$ 0.364} & -20.68 \textr{$\pm$ 0.435} & -20.55 \textr{$\pm$ 0.120} & -21.93 \textr{$\pm$ 0.112} & \textbf{-23.82} \textr{$\pm$ 0.270} \\
\hline
\multirow{2}{*}{\rotatebox{90}{\scriptsize $R=3$}}
& SSIM & 0.930 \textr{$\pm$ 0.0074} & 0.969 \textr{$\pm$ 0.0036} & 0.898 \textr{$\pm$ 0.0051} & 0.939 \textr{$\pm$ 0.0059} & \textbf{0.973} \textr{$\pm$ 0.0021} \\
& NMSE & -19.23 \textr{$\pm$ 0.228} & -19.21 \textr{$\pm$ 0.276} & -16.87 \textr{$\pm$ 0.103} & -18.91 \textr{$\pm$ 0.080} & \textbf{-22.91} \textr{$\pm$ 0.218}\\
\hline
\multirow{2}{*}{\rotatebox{90}{\scriptsize $R=4$}}
& SSIM & 0.897 \textr{$\pm$ 0.0088} & 0.957 \textr{$\pm$ 0.0058} & 0.837 \textr{$\pm$ 0.0057} & 0.909 \textr{$\pm$ 0.0083} & \textbf{0.967} \textr{$\pm$ 0.0021} \\
& NMSE & -17.21 \textr{$\pm$ 0.181} & -18.50 \textr{$\pm$ 0.403} & -14.45 \textr{$\pm$ 0.094} & -16.99 \textr{$\pm$ 0.079} & \textbf{-21.62} \textr{$\pm$ 0.229}\\
\hline
\multirow{2}{*}{\rotatebox{90}{\scriptsize $R=5$}}
& SSIM & 0.849 \textr{$\pm$ 0.0112} & 0.945 \textr{$\pm$ 0.0080} & 0.771 \textr{$\pm$ 0.0040} & 0.873 \textr{$\pm$ 0.0106} & \textbf{0.962} \textr{$\pm$ 0.0029} \\
& NMSE & -14.96 \textr{$\pm$ 0.188} & -17.77 \textr{$\pm$ 0.449} & -12.62 \textr{$\pm$ 0.159} & -15.40 \textr{$\pm$ 0.062} & \textbf{-21.17} \textr{$\pm$ 0.140}\\ 
\hline
\multirow{2}{*}{\textr{\rotatebox{90}{\scriptsize $R=6$}}} 
& \textr{SSIM} & \textr{0.789} \textr{$\pm$ 0.0120} & \textr{0.925} \textr{$\pm$ 0.0095} & \textr{0.758} \textr{$\pm$ 0.0043} & \textr{0.853} \textr{$\pm$ 0.0113} & \textbf{\textr{0.943}} \textr{$\pm$ 0.0037} \\
& \textr{NMSE} & \textr{-12.01} \textr{$\pm$ 0.331} & \textr{-17.17} \textr{$\pm$ 0.389} & \textr{-11.12} \textr{$\pm$ 0.105} & \textr{-13.70} \textr{$\pm$ 0.103} & \textbf{\textr{-20.278}} \textr{$\pm$ 0.239}\\
\hline
\end{tabular}
\end{adjustbox}
\caption{\textr{Quantitative results from the retrospective patient LGE (Study III). The rows show NMSE and SSIM values at $R=2, 3, 4,5,$ and $6$. Average values ($\pm$SEM) are reported from eight distinct patient series. Bold values indicate the best results.}}
\label{tab:LGE_retro_tab}
\end{table*}

\tabref{LGE_retro_tab} compares \discus with CS, L+S, DIP, and SG-DIP at four different acceleration rates of $R=2,~3,~4,~\text{and}~5$. Consistent with the previous simulation study, \discus offers a significant advantage over other methods at all four acceleration rates, while L+S marginally outperforms \discus in terms of SSIM at $R=2$. \figref{LGE_retro_fig} shows a representative frame from one of the image series at $R=4$. In this example, \discus suppresses noise and preserves fine details, while CS shows excessive blocky artifacts, and L+S shows an artifact as highlighted by red arrow. Both DIP and SG-DIP exhibit excessive noise amplification. \figref{LGE_retro_R5_fig} shows a frame from the same patient but at a higher acceleration ($R=5$); here, CS and DIP reconstructions show significant quality degradation.

\subsection{Study IV--LGE with Prospective Undersampling}

\begin{figure*}[!h]
  \centering
  \begin{tikzpicture}[scale=1]
    \node[anchor=north west, inner sep=0] (image) at (0,0) 
    {\includegraphics[width=0.81\textwidth]{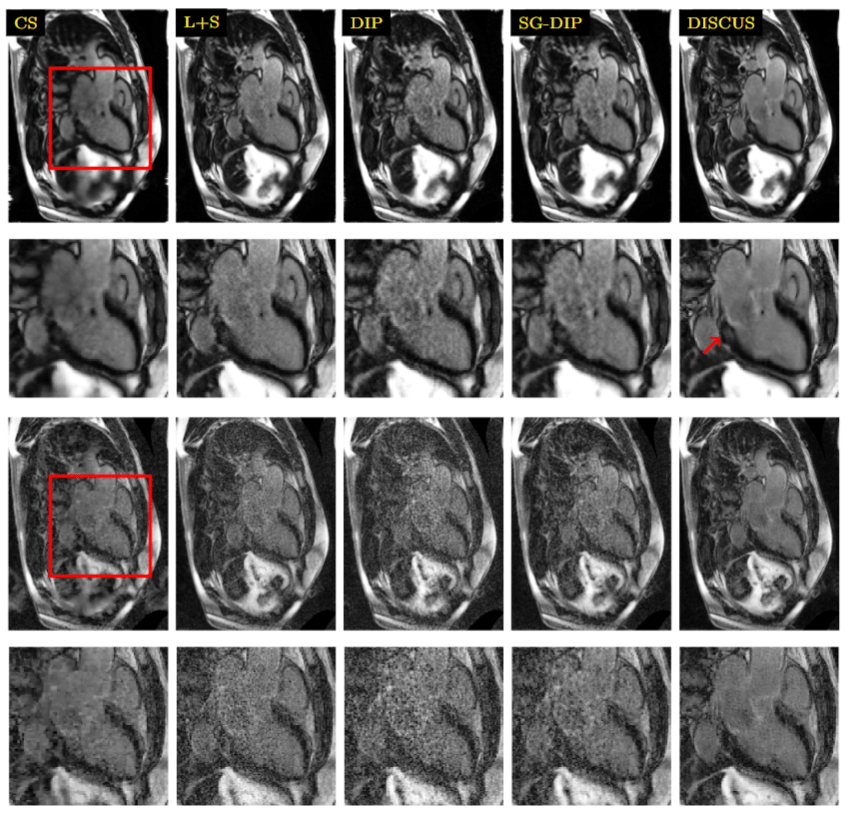}};

  \end{tikzpicture}
  \caption{\textb{Representative results from the prospective undersampled patient LGE (Study IV) at $R=4$ with and without motion correction. First row shows an example motion-corrected frame from one of the eight patients with reconstructions by CS, L+S, DIP, SG-DIP, and \discus. The second row provides a zoomed-in view of the red box in first row. The enhancement in the basal inferolateral wall is more visible in \discus as highlighted by the red arrow. Third row shows an example frame without motion correction from the same patient. The final row provides a zoomed-in view of the red box in third row. The advantage of \discus is more evident in last row without motion correction.}}
  \label{fig:LGE_pros_fig}
\end{figure*}

\begin{table}[!h]
\centering
\renewcommand{\arraystretch}{1.85}
\begin{adjustbox}{width=0.475\textwidth}
\fontsize{14}{14}\selectfont
\begin{tabular}{|c|ccccc|}
\hline
& \textbf{CS} &\textbf{L+S} &\textbf{DIP} &\textbf{SG-DIP} &\textbf{\discus} \\

\hline
Avg.& 2.5 \textr{$\pm$ 0.13} & 4.5 \textr{$\pm$ 0.16} & 3.75 \textr{$\pm$ 0.17} & 4.1 \textr{$\pm$ 0.17} & \textbf{4.63} \textr{$\pm$ 0.13} \\
\hline
Count& 0 & 6  & 1 & 2  & \textbf{11} \\
\hline
\end{tabular}
\end{adjustbox}
\caption{\textr{Quantitative results for the prospective undersampled LGE images (Study IV). The first row represents an average score ($\pm$SEM) from two cardiac MRI experts for 8 datasets, reviewed in a blinded manner. The last row shows the number of times the method was selected by a reviewer as the best reconstruction for a given dataset. The bold value indicates the best scores.}}
\label{tab:LGE_pros_tab}
\end{table}



\tabref{LGE_pros_tab} compares \discus with CS, L+S, DIP, and SG-DIP in terms of expert scoring. In terms of the average score, \discus outperforms CS, DIP, and SG-DIP comprehensively but offers a minor advantage over L+S. We attribute the seemingly diminished advantage of \discus over other methods to the use of MoCo, which narrows the separation between different methods by introducing additional blurring. In terms of best image count, \discus substantially outperforms other methods.\textb{A representative LGE image at $R=4$ is shown in \figref{LGE_pros_fig}, both with and without MoCo. Compared to other methods, \discus preserves the fine details, including enhancement in the basal inferolateral wall (red arrow), without noise amplification in second row with MoCo. The last two rows show an individual frame from the same image series. Without MoCo, the advantage of \discus becomes even more apparent, as it effectively suppresses noise while maintaining image quality.}

\section{Discussion}\label{sec:dis}

Training data are not readily available for dynamic MRI applications. In this paper, we proposed a DIP-inspired unsupervised method to jointly reconstruct a series of images. DISCUS does not require pre-specifying the dimensionality of the manifold and discovers it by enforcing group sparsity on the code vectors. Additionally, it does not enforce smooth transitions between the neighboring frames. In the first two studies, we validated DISCUS on simulated phantoms. In studies III and IV, we applied DISCUS on measured LGE data.

In the first simulated study, we evaluated \discus's ability to discover the true manifold dimensionality by enforcing group sparsity on dynamic code vectors. For image series with only rotation or translation, where the true dimensionality was one, \discus correctly identified it in all 20 instances. 
For series involving both motions, \discus discovered the correct dimensionality of two in eight out of ten cases; in the remaining two, it identified the dimensionality to be three. However, in those two cases with incorrect dimensionality, NMSE values were within 0.5 dB of the average value. Even for the eight cases where \discus correctly discovered the manifold dimensionality, disentanglement between rotation and translations was not observed. In other words, manipulating the two non-zero entries in $\hvec{z}_t$ separately did not always lead to pure rotations or translations of the output images. This disentanglement may not be needed to generate high-quality images but may have value in separating different physiological motions. Further efforts are needed to improve the disentanglement of different components of the manifold.


In the second simulated study, we compared DISCUS with CS, L+S, DIP, and SG-DIP at multiple accelerations using a realistic LGE phantom. We found out that DISCUS outperforms the other methods across different acceleration rates in terms of both NMSE and SSIM. We also performed ablations to assess the contribution of group sparsity and the impact of the number of frames. As expected, L+S, DGS, and \discus benefit from the availability of a larger number of frames, with \discus outperforming L+S and DGS at all acceleration rates and preferentially benefiting from the larger value of $T$. We also observed that DGS, which does not employ group sparsity, is prone to overfitting and did not perform well when the number of iterations was increased beyond 10,000. In contrast, \discus offered a flatter convergence curve, with NSME values not changing significantly between 8,000 and 15,000 iterations.


In Studies III and IV, \discus was again compared to CS, L+S, DIP, and SG-DIP. For the retrospectively sampled study, where the ground truth was available, \discus clearly outperformed the competing methods. This was evident both in the NMSE and SSIM numbers in \tabref{LGE_retro_tab} and error maps in \figref{LGE_retro_fig}. For the prospectively undersampled data, \discus images were considered the best 11 times, compared to six times for L+S. However, in terms of average score, \discus outperformed the second-best method (L+S) only marginally. We believe that this is due to the subjective nature of the scoring and the extra layer of processing introduced by MoCo. The images were scored after MoCo, because cardiologists at our institution typically read single-shot LGE images that have been motion corrected. Although MoCo suppresses noise, it also introduces blurring due to imperfect registration and through-plane motion. As shown in \figref{LGE_pros_fig}, the advantage of \discus over other methods is more evident before MoCo. \textr{Future efforts will focus on extending \discus to other single-shot applications, such as first-pass perfusion and tissue mapping, and accelerate \discus training to reduce reconstruction times.}

\section{Conclusions}\label{sec:con}
In this paper, we proposed an unsupervised method, called \discus, that does not rely on fully sampled data for training. Using simulation and measured data, the performance of \discus is evaluated for single-shot LGE imaging. \discus outperforms other unsupervised or self-supervised methods in terms of image quality and expert reader scoring.

\section*{Statements and Declarations}
\textbf{Supplementary Information} The following supporting information is available as part of the online article: ``ESM\_1.mp4''.

\vspace{2mm}
\noindent
\textbf{Funding} This work was partially supported by NIH Grants R01-EB029957 and R01-HL151697.

\vspace{2mm}
\noindent
\textbf{Acknowledgments} The authors would like to thank Xuan Lei for assisting with MRXCAT simulation. 
\textb{The authors would also like to thank Siemens Healthineers for providing MoCo code.}

\vspace{2mm}
\noindent
\textbf{Ethics approval} For the human subject data, approval was granted by the Institutional Review Board (IRB) at The Ohio State University (2019H0076).

\vspace{2mm}
\noindent
\textbf{Consent to participate} Informed consent was obtained from all individual participants included in this work.

\vspace{2mm}
\noindent
\textbf{Consent to publish} Consent to publish and disseminate results were received from all human participants.

\vspace{2mm}
\noindent
\textbf{Competing interests} The authors declare no competing interests.

\vspace{2mm}
\noindent
\textbf{Conflict of interest} The authors declare no conflict of interest.

\vspace{2mm}
\noindent
\textbf{Code and Data availability} \discus implementation and representative datasets are available on GitHub at \url{https://github.com/OSU-MR/discus}.

\vspace{2mm}
\noindent
\textbf{Authors' contributions} M.A. Sultan coded \discus and prepared the first draft, C. Chen and Y. Liu assisted with data acquisition and pre-processing, K. Gil and K. Zareba reviewed the images and provided clinical expertise, and R. Ahmad advised the first author and co-authored the manuscript.

\clearpage

\bibliography{root}%

\clearpage 

\end{document}